\documentclass[12pt]{article}
\usepackage{fullpage,epsfig,fancyheadings}
\usepackage[psamsfonts]{euscript}
\usepackage{epic}
\usepackage{amstex}
\numberwithin{equation}{section} 
\newcommand{\beq}{\begin{equation}}
\newcommand{\eeq}{\end{equation}}

\begin{document}
\vspace*{-.6in}
\thispagestyle{empty}
\begin{flushright}
CALT-68-2158\\
DOE RESEARCH AND\\
DEVELOPMENT REPORT
\end{flushright}
\baselineskip = 20pt

\vspace{.5in}
{\Large
\begin{center}
HQET and Exclusive $B$ Decays\footnote{Work supported in part by the U.S. Dept.
of Energy under Grant No. DE-FG03-92-ER40701.}
\end{center}}
\vspace{.4in}

\begin{center}
Mark B. Wise \\
\emph{California Institute of Technology, Pasadena, CA  91125 USA}
\end{center}
\vspace{0.5in}
\begin{center}
Talk delivered at Beauty 97\\
University of California, Los Angeles \\
October 12-16, 1997.
\end{center}
\vspace{0.5in}

\begin{center}
\textbf{Abstract}
\end{center}
\begin{quotation}
\noindent  Exclusive semileptonic $B$ decays are discussed.  The emphasis is on
using semileptonic decays to determine $|V_{ub}|$ and $|V_{cb}|$.  Recent
progress in our understanding of $B$ semileptonic decays to excited charmed
mesons is also reviewed.
\end{quotation}

\newpage


\section{Introduction}

In the area of exclusive $B$ decay most applications of heavy quark effective
theory (HQET) have been to
semileptonic decays, which is the topic I will concentrate on in this
lecture.  In Section II, I review the determination of $|V_{cb}|$ from exclusive
$B\rightarrow D^* \ell \bar \nu_{\ell}$ decay.  Section III reviews recent
theoretical
progress in our understanding of $B$ semileptonic decay to the doublet of
excited charmed mesons with spin of the light degrees of freedom, $s_\ell =
{3\over2}$, and positive parity.  Finally, in Section IV prospects for
determining
$|V_{ub}|$ from data on exclusive semileptonic $B$ (and $D$) decays are
discussed. Other promising methods for determining $|V_{cb}|$ and $|V_{ub}|$
from $B$ decays involve using the operator product expansion (OPE) to
predict inclusive decay rates and lattice QCD to predict exclusive decay matrix
elements.
However, these techniques will not be discussed in this lecture.

\section{$|V_{cb}|$ and Exclusive $B \rightarrow D^* \ell\bar \nu_\ell$ Decay}

In the near future much of the high energy experimental program will be devoted
to testing whether the KM phase~\cite{kobayashi} is responsible for the CP
violation we observe
in nature.  This program revolves around using $B$-decays to determine the
sides and angles of the unitarity triangle in as many ways as possible and
checking for inconsistencies in the results.
Since the deviations from the standard model
may not be large, a precise determination of the angles and
sides of the triangle is desirable.

At the present time CP nonconservation has only been
observed in kaon decays. In the standard
model it arises from second order weak $K^0 - \bar K^0$ mixing.
The CKM elements that
occur in the box diagram with a top quark in the loop are $(V^*_{td} V_{ts})^2
=
(\rho - 1 - i\eta )^2 |V_{cb}|^4|V_{us}|^2$.  (With the usual conventions $\rho + i\eta$
 are the coordinates in the complex plane of the vertex of the unitarity
triangle that doesn't lie on the real axis.)  Therefore, $|V_{cb}|$ must be
known very
accurately if the measured value of the CP violation parameter $\varepsilon$ is
to be compared with theory.

The semileptonic form factors that occur in $B \rightarrow D^{(*)}
e \bar\nu_e$ decay are defined by
\begin{align}
{\langle D(v')|\bar c\gamma^{\mu} b| B(v)\rangle\over\sqrt{m_D m_B}} &= h_+ (w)
(v + v')^\mu + h_ - (w) (v - v')^\mu, \\
{\langle D^*(v',\varepsilon)|\bar c\gamma^{\mu} b|
B(v)\rangle\over\sqrt{m_{D^{*}} m_B}} &= ih_V (w)
\varepsilon^{\mu\nu\alpha\beta} \varepsilon^*_\nu v'_\alpha v_\beta, \\
{\langle D^*(v',\varepsilon)|\bar c\gamma^{\mu} \gamma_5
b|B(v)\rangle\over\sqrt{m_{D^{*}} m_B}} &= h_{A_{1}} (w) (w + 1)
\varepsilon^{*\mu} - h_{A_{2}} (w) (\varepsilon^* \cdot v) v^\mu - h_{A_{3}}
(w) (\varepsilon^* \cdot v) v^{\prime \mu}.
\end{align}
In eqs. (2.1), (2.2) and (2.3) the $h_j$ are Lorentz scalar form factors that are
functions of the dot product of the $B$ and $D^*$ four-velocities $w = v \cdot
v'$.  In the rest frame of the $B$ meson $w = E_{D^{(*)}}/m_{D^{(*)}}$ is the
$\gamma$ factor for the recoiling $D^{(*)}$.  At zero recoil $w = 1$.

Heavy quark spin symmetry implies that for $m_{c,b} \rightarrow \infty$ the
form
factors are given by~\cite{isgur}: $h_-(w)
= h_{A_{2}} (w) = 0$ and $h_+ (w) = h_V (w) = h_{A_{1}} (w) = h_{A_{3}} (w) =
\xi (w)$.  At the kinematic point $w = 1$ the QCD operator $\bar c \gamma_\mu
b$ matches onto  a generator of the flavor-spin symmetry in
HQET~\cite{eichten1}.  Since the
matrix elements of the symmetry generators are known, heavy quark
symmetry implies the normalization condition~\cite{isgur,nussinov}
\begin{equation}
\xi (1) = 1.
\end{equation}
There are perturbative corrections to these predictions suppressed by $\alpha_s
(m_{c,b})/\pi$~\cite{falk1,falk2,neubert} and nonperturbative corrections
suppressed by
$\Lambda_{QCD}/m_{c,b}$~\cite{eichten2,luke}.  The perturbative corrections are
calculable and do
not cause a loss of predictive power.

The differential decay rate for $B \rightarrow D^* e\bar \nu_e$ is
\begin{eqnarray}
{d\Gamma\over dw} (B \rightarrow D^* \ell\bar\nu_e) &=& {G_F^2 m_B^5\over
192\pi^3} r^3 (1 - r)^2 (w^2 - 1)^{1/2} (w + 1)^2 \nonumber \\ 
&&\times \left[ 1 + {4w\over w + 1} {1 - 2wr + r^2\over (1 - r)^2}\right]
|V_{cb}|^2 |
{\cal F}_{B \rightarrow D^{*}} (w)|^2.
\end{eqnarray}
where $r=m_{D^*}/m_B$.
${\cal F}_{B\rightarrow D^{*}} (w)$ can be expressed in terms of the form
factors $h_j(w)$ and in the $m_{c,b} \rightarrow \infty$ limit ${\cal
F}_{B\rightarrow D^{*}}
(w) = \xi(w)$.  The known normalization at $w = 1$ allows an
extraction of $|V_{cb}|$ from an extrapolation of data on this decay to
the zero recoil kinematic point.

The structure of the symmetry breaking corrections to ${\cal F}_{B\rightarrow
D^{*}} (1)$ is
\begin{equation}
{\cal F}_{B\rightarrow D^{*}} (1) = 1 + \delta\eta_A (\alpha_s) + 0 +
\delta_{1/m_{c,b}^{2}} + \ldots = 0.91 \pm 0.05.
\end{equation}
In the above $\delta\eta_A(\alpha_s)$ is the perturbative QCD correction.  It
is known to order $\alpha_s^2$ and has the value $-0.04$
~\cite{czarnecki}.  There
is no correction
of order $\Lambda_{QCD}/m_{c,b}$ and hence the zero for the third term in eq.
(2.6).  This result is known as Luke's theorem~\cite{luke}.  The terms of order
$(\Lambda_{QCD}/m_{c,b})^2$ and higher are estimated by
models~\cite{falkneubert,randallwise,mannel} to have the
value $-0.05$.  Since this is a model dependent contribution I assign it a 100\%
theoretical uncertainty and this is the source of the theoretical error on the
right hand side of eq. (2.6).  The experimental data gives~\cite{drell,gibbons}
$|V_{cb} {\cal
F}_{D^{*}} (1)| = (35.2 \pm 1.4) \times 10^{-3}$ which when combined with eq.
(2.6) yields
\begin{equation}
|V_{cb}| = (38.6 \pm 2.3_{exp} \pm 2_{th}) \times 10^{-3}.
\end{equation}

Part of the nonperturbative corrections to ${\cal F}_{B \rightarrow D^{*}}(1)$
are calculable in a model independent way.  This part has a nonanalytic
dependence on the light quark masses and takes the form~\cite{randallwise}
\begin{equation}
\delta_{1/m_{c}^{2}} + \ldots = {g^2 \Delta^2\over (4\pi f_\pi)^2}
Y(\Delta/m_\pi),
\end{equation}
where $Y$ is a known function of $\Delta = m_{D^*} - m_{D}$ divided by the pion
mass. (For simplicity of presentation in eq. (2.8) I have taken the limit $m_b
\rightarrow \infty$ and only kept nonperturbative corrections suppressed by
powers of $1/m_c$.)  In eq. (2.8) $g$ is the $D^* D\pi$ coupling and $f_\pi$
is the pion decay constant.  The coupling $g$ also occurs in the $D^*$ decay
width
\begin{equation}
\Gamma (D^{*+} \rightarrow D^0 \pi^+) = {1\over 6\pi} {g^2\over f_\pi^2} |\vec
p_\pi|^3 \simeq 0.2 g^2~{\rm MeV}.
\end{equation}
The branching ratio for $D^{*+} \rightarrow D^0 \pi^+$ is 68.3\%.  A
measurement of the $D^*$
width would fix $g$ and reduce somewhat the theoretical
uncertainty in
eq. (2.6).  Our expectation, based on the chiral quark model
{}~\cite{georgimanohar} is that $g^2 \sim
1/2$, but the uncertainty in this estimate is very large.

Part of the experimental error in the determination of $|V_{cb}|$ arises from
the extrapolation to zero recoil. This can be reduced by using a parametrization
for the $w$ dependence of ${\cal F}_{B \rightarrow D^{*}}(w)$ that is 
 constrained by dispersion relations and perturbative QCD~\cite{grinsteinboyd,
neubert1}.

The error estimate associated with theory in eq. (2.7) is rather adhoc.  Of
course there is not really a correct way to assign an error from
theory.  What one needs is another method for determining $|V_{cb}|$.
The consistency between it and the exclusive method then
provides a measure of the theoretical uncertainties.  Fortunately
such a method exists which uses the inclusive $B$ semileptonic decay
rate~\cite{drell}.  At
the present time this approach has a somewhat smaller experimental uncertainty.
The theoretical uncertainty associated with the inclusive technique is also
expected to be small. However, uncertanties from
 possible violations of quark-hadron duality that are not apparent at low
orders in the OPE are very difficult to estimate.
The inclusive way of determining $|V_{cb}|$ yields a value close to that in
eq. (2.7) indicating that theoretical uncertainties in both of these methods
are indeed less than 5\% .

\section{$B$ Semileptonic decay to Excited Charm Mesons}

The members of the excited $s_\ell^\pi = {3\over 2}^+$ doublet have been
observed.
They are the $D_1 (2420)$ and $D_2^* (2460)$.  Recently there have been
measurements of the $B$ branching ratio to $D_1 e\bar\nu_e$ and a limit on the
branching ratio to the $D_2^* e\bar\nu_e$ final state~\cite{aleph,cleo},
\begin{align}
Br (B^- \rightarrow D_1^0 e^- \bar\nu_e) &= (0.56 \pm 0.14)\%,\\
Br (B^- \rightarrow D_2^{*0} e^- \bar\nu_e) &\leq 0.8\%.
\end{align}
The branching ratios to the ground state doublet, $Br(B \rightarrow
De\bar\nu_e) = 1.8 \pm 0.4\%$ and $Br (B \rightarrow D^* e\bar\nu_e) = 4.6 \pm
0.3\%$, indicate that about $35\%$ of $B$ semileptonic decays are to
excited mesons and nonresonant final states.

In terms of Lorentz scalar form factors the matrix elements of the weak vector
and axial vector form factors are
\begin{align}
{\langle D_1 (v',\varepsilon) | \bar c \gamma^\mu b|B(v)\rangle\over
\sqrt{m_{D_{1}}m_B}} &= f_{V_{1}}\varepsilon^{*\mu} + (f_{V_{2}} v^\mu +
f_{V_{3}} v^{\prime\mu})(\varepsilon^* \cdot v),\\
{\langle D_1 (v',\varepsilon) | \bar c \gamma^\mu \gamma_5 b|B(v)\rangle\over
\sqrt{m_{D_{1}} m_B}} &= i f_A \varepsilon^{\mu\alpha\beta\gamma}
\varepsilon_\alpha^* v_\beta v'_\gamma,\\
{\langle D_2^* (v',\varepsilon) | \bar c \gamma^\mu \gamma_5 b|B(v)\rangle\over
\sqrt{m_{D_{2}^{*}} m_B}} &= k_{A_{1}} \varepsilon^{*\mu\alpha} v_\alpha +
(k_{A_{2}} v^\mu + k_{A_{3}} v^{\prime\mu}) \varepsilon_{\alpha\beta}^*
v^\alpha v^{\beta},\\
{\langle D_2^* (v',\varepsilon) | \bar c \gamma^\mu \gamma_5 b|B(v)\rangle\over
\sqrt{m_{D_{2}^{*}} m_B}} & = i k_V \varepsilon^{\mu\alpha\beta\gamma}
\varepsilon_{\alpha\eta}^* v^\eta v_\beta v'_\gamma.
\end{align}
The form factors $f_j$ and $k_j$ are functions of $w$.  Note that the $B
\rightarrow D_2^*$ zero recoil matrix elements of the vector and axial vector
currents vanish by Lorentz invariance, independent of the values of $k_j(1)$.
However, the $B \rightarrow D_1$ zero recoil matrix element of the vector
current is nonzero if $f_{V_{1}} (1) \not= 0$.  Heavy quark symmetry implies
that in the $m_{c,b} \rightarrow \infty$ limit, $f_{V_1}(1) = 0$.  Since most
of
the phase space is near zero
recoil, $1 <w <1.3$, the $\Lambda_{QCD}/m_{c,b}$ 
corrections which cause the zero
recoil $D_1$ matrix element
not to vanish are very important.

In the limit $m_{c,b}\rightarrow \infty$ heavy quark spin symmetry implies that
the form factors $f_j$ and $k_j$ can be expressed in terms of a single function
of $w$~\cite{isgur2}.
\begin{equation}
\begin{array}{ll}
\sqrt{6} f_A (w) = - (w + 1) \tau (w) \qquad \qquad &k_V (w) = - \tau
(w)\\
\sqrt{6} f_{V_{1}} (w) =  (1 - w^2) \tau (w) \qquad \qquad &k_{A_{1}} (w) = -
(1 + w)\tau (w) \\
\sqrt{6} f_{V_{2}} (w) = - 3\tau (w) \qquad \qquad &k_{A_{2}} (w) =  0 \\
\sqrt{6} f_{V_{3}} (w) =  (w - 2) \tau (w) \qquad \qquad &k_{A_{3}} (w) =  \tau
(w)\end{array}
\end{equation}
Note that unlike the $B\rightarrow D^{(*)} e\bar\nu_e$ case $\tau(1)$ is not
fixed by heavy quark symmetry.  In the infinite mass limit $f_{V_{1}} (1) =
0$ because of the factor of $(1 - w^2)$ in eq. (3.7).

An analysis of the $\Lambda_{QCD}/m_{c,b}$ corrections gives~\cite{leibovich}
\begin{equation}
\sqrt{6} f_{V_1} = - 4 (\bar\Lambda^* - \bar\Lambda) \tau (1)/m_c,
\end{equation}
where $\bar\Lambda$ is the mass of the light degrees of freedom in a member of
the $s_\ell^\pi = {1\over 2}^{-}$ doublet and $\bar\Lambda^*$ is the mass of
the
light degrees of freedom in a member of the $s_\ell^\pi = {3\over 2}^+$
doublet.  The
difference $\bar\Lambda^* - \bar\Lambda$ can be expressed in terms of known
hadron masses yielding $\bar\Lambda^* - \Lambda \simeq~ 0.39 {\rm GeV}$.  This
is the
most important $\Lambda_{QCD}/m_{c,b}$ correction because it is the only one at
zero recoil.

Away from zero recoil other $\Lambda_{QCD}/m_{c,b}$ corrections arise and some
model dependence occurs.  In the infinite mass limit $R = Br (B \rightarrow
D_2^* e\bar\nu_e)/Br (B\rightarrow D_1 e\bar\nu_e) = 1.65$ and the measured
value of the $B\rightarrow D_1 e\bar\nu_e$ decay rate implies $ |\tau(1)| =
1.27$.  Including the $\Lambda_{QCD}/m_{c,b}$ corrections changes these results
to~\cite{leibovich} (see also ~\cite{neubert2}) $R
\simeq 0.6$ and $|\tau(1)|\simeq 0.7$.  (For these predictions $
\tau'(1) / \tau(1) = - 1.5$ and $\tau(w) = \tau(1) +\tau '(1) (w - 1)$
were used.)  The $\Lambda_{QCD}/m_{c,b}$ corrections lead to the expectation
that $R < 1$, which is opposite from what the infinite mass limit gives.

\section{$|V_{ub}|$ From Exclusive $B$ Decay}

Recently branching ratios for $B \rightarrow \pi e\bar\nu_e$ and
$B\rightarrow\rho e\bar\nu_e$ have been measured~\cite{alexander}.
  One of the original
applications of heavy quark symmetry was to take the measured $D \rightarrow
K^* \bar e\nu_e$ form factors and use $SU(3)$ (light quark) flavor symmetry and
heavy quark symmetry to get $B \rightarrow \rho e\bar\nu_e$ form
factors~\cite{isgur3}.  For such decays the
form factors are defined by
\begin{align}
\langle V(p', \varepsilon) | \bar q \gamma_\mu Q| H(p)\rangle &=
ig^{(H\rightarrow V)} \varepsilon_{\mu\nu\lambda\sigma}\varepsilon^{*\nu} (p + p')^\lambda (p -
p')^\sigma, \\
\langle V(p', \varepsilon) | \bar q \gamma_\mu \gamma_5 Q|H(p)\rangle &=
f^{(H\rightarrow V)} \varepsilon_\mu^* + a_+^{(H\rightarrow V)} (\varepsilon^*
 \cdot p) (p + p')_\mu + a_-^{(H \rightarrow V)} (\varepsilon^* \cdot p) (p -
p')_\mu. \notag
\end{align}
It is convenient to view the form factors $f,g$ and $a_\pm$ as functions of $y
= v
\cdot v'$ where $p = m_H v$ and $p' = m_V v'$.  Then $q^2 = m_H^2 + m_V^2 - 2
m_H m_V y$. In this section $y$ is used for $v\cdot v^{\prime}$ (instead of
$w$) as a reminder that the light up down and strange quarks are not treated
 as heavy. Assuming pole dominance for the form factors the $D\rightarrow K^*
\bar e \nu_e$ data gives~\cite{aitala1}
\begin{align}
f^{(D\rightarrow K^{*})} (y) &= {(1.9 \pm 0.1) {\rm GeV}\over 1 + 0.63 (y -
1)},\notag
\\
a_+^{(D \rightarrow K^{*})} (y) &= - {(0.18 \pm 0.03) {\rm GeV}^{-1}\over 1 +
0.63 (y
- 1)},\notag \\
g^{(D\rightarrow K^{*})} (y) &=- {(0.49 \pm 0.04) {\rm GeV}^{-1}\over 1+ 0.96
(y -
1)}.
\end{align}
These form factors are measured only over the kinematic region $1 < y < 1.3$.
Over
this range $f^{(D \rightarrow K^{*})}$ changes by less than 20\%.  However, the
full kinematic range for $B \rightarrow \rho e\bar\nu_e$ is much larger, $1 < y
<
3.5$.

The differential decay rate for $B \rightarrow \rho e\bar\nu_e$ is
\begin{equation}
{d\Gamma (B \rightarrow \rho e\bar\nu_e)\over dy} = {G_F^2 |V_{ub}|^2\over
48\pi^3} m_B m_{\rho}^2 S^{(B\rightarrow \rho)} (y),
\end{equation}
where
\begin{eqnarray}\label{shape}
S^{(H\to V)}(y) &=& \sqrt{y^2-1}\, \bigg[ \Big|f^{(H\to V)}(y)\Big|^2\,
  (2+y^2-6yr+3r^2) \nonumber\\*
&&\phantom{} + 4{\rm Re} \Big[a_+^{(H\to V)}(y)\, f^{(H\to V)*}(y)\Big]
  m_H^2\, r\, (y-r) (y^2-1) \nonumber\\*
&&\phantom{} + 4\Big|a_+^{(H\to V)}(y)\Big|^2 m_H^4\, r^2 (y^2-1)^2 +
  8\Big|g^{(H\to V)}(y)\Big|^2 m_H^4\, r^2 (1+r^2-2yr)(y^2-1)\, \bigg]
  \nonumber\\*
&=& \sqrt{y^2-1}\, \Big|f^{(H\to V)}(y)\Big|^2\, (2+y^2-6yr+3r^2)\,
  [1+\delta^{(H\to V)}(y)] \,,
\end{eqnarray}
with $r=m_V/m_H$.  The function $\delta^{(H\to V)}$ depends on the ratios of
form factors $a_+^{(H\to V)}/f^{(H\to V)}$ and $g^{(H\to V)}/f^{(H\to V)}$.
$S^{(B\to\rho)}(y)$ can be estimated using combinations of $SU(3)$ flavor
symmetry and heavy quark symmetry.  $SU(3)$ symmetry implies that the $
B^0\to\rho^+$ form factors are equal to the $B\to K^*$ form factors and the
$B^-\to\rho^0$ form factors are equal to $1/\sqrt2$ times the $B\to K^*$ form
factors.  Heavy quark symmetry implies the relations
\begin{eqnarray}\label{BDrel}
f^{(B\to K^*)}(y) &=& \left({m_B\over m_D}\right)^{1/2}
  f^{(D\to K^*)}(y)\,, \nonumber\\*
a_+^{(B\to K^*)}(y) &=& \left({m_D\over m_B}\right)^{1/2}
  a_+^{(D\to K^*)}(y) \,, \nonumber\\*
g^{(B\to K^*)}(y) &=& \left({m_D\over m_B}\right)^{1/2}
  g^{(D\to K^*)}(y)\,.
\end{eqnarray}
The second relation is obtained using $a_-^{(D\to K^*)}=-a_+^{(D\to K^*)}$,
valid in the large $m_c$ limit.

Using eq.~(\ref{BDrel}) and $SU(3)$ to get the
$B^0\to\rho^+\,\ell\,\bar\nu_\ell$ form factors (in the region $1<y<1.5$) from
those for $D\to K^*\bar\ell\,\nu_\ell$ given in eq. (19) yields
$S^{(B\to\rho)}(y)$ plotted in Fig.~1 of Ref.~\cite{ligeti1}.  The numerical
values
in eq. (19) differ slightly from those used in Ref.~\cite{ligeti1}.  This
makes only a small difference in $S^{(B\to\rho)}$, but changes
$\delta^{(B\to\rho)}$ more significantly.  In the region $1<y<1.5$,
$|\delta^{(B\to\rho)}(y)|$ defined in eq.~(\ref{shape}) is less than 0.06,
indicating that $a_+^{(B\to\rho)}$ and $g^{(B\to\rho)}$ make a small
contribution to the differential rate in this region.

This prediction for $S^{(B\to\rho)}$ can be used to determine $|V_{ub}|$ from a
measurement of the $B\to\rho\,\ell\,\bar\nu_\ell$ semileptonic decay rate in
the region $1<y<1.5$.  This method is model independent, but cannot be expected
to yield a very accurate value of $|V_{ub}|$.  Typical $SU(3)$ violations are
at the $10-20$\% level and one expects similar violations of heavy quark
symmetry.

Ref.~\cite{ligeti1} proposed a method for getting a value of
$S^{(B\to\rho)}(y)$
with small theoretical uncertainty.  They noted that the
``Grinstein-type"~\cite{grinstein}  double ratio
\begin{equation}\label{Gtdr}
R(y) = \Big[ f^{(B\to\rho)}(y) / f^{(B\to K^*)}(y) \Big] \Big/
  \Big[ f^{(D\to\rho)}(y) / f^{(D\to K^*)}(y) \Big]
\end{equation}
is unity in the limit of $SU(3)$ symmetry or in the limit of heavy quark
symmetry.  Corrections to the prediction $R(y)=1$ are therefore suppressed by
$m_s/m_{c,b}$ ($m_{u,d} \ll m_s$) instead of $m_s/\Lambda_{\rm QCD}$ or
$\Lambda_{\rm QCD}/m_{c,b}$.  A recent estimate of
$R(1)$ using chiral perturbation theory gives~\cite{ligeti2} $R(1) = 1 -
0.035 gg_2$ where
$g_2$ is the $\rho \omega \pi$ coupling. Experimental information
 on $\tau \rightarrow \omega \pi \nu_{\tau}$ decay yields $g_2 \simeq 0.6$
{}~\cite{hooman}.

Since $R(y)$ is very close to unity, the
relation
\begin{equation}\label{magic}
S^{(B\to\rho)}(y) = S^{(B\to K^*)}(y)\,
  \left|{f^{(D\to\rho)}(y)\over f^{(D\to K^*)}(y)}\right|^2\,
  \bigg({m_B-m_\rho\over m_B-m_{K^*}}\bigg)^2\,,
\end{equation}
together with measurements of $|f^{(D\to K^*)}|$, $|f^{(D\to\rho)}|$, and
$S^{(B\to K^*)}$ will determine $S^{(B\to\rho)}$ with small theoretical
uncertainty.  The last term on the right-hand-side makes eq.~(\ref{magic})
equivalent to eq.~(\ref{Gtdr}) in the $y\to1$ limit.  The ratio of the
$(2+y^2-6yr+3r^2)\,[1+\delta^{(B\to V)}(y)]$ terms makes only a small and
almost $y$-independent contribution to $S^{(B\to\rho)}/S^{(B\to K^*)}$ in the
range $1<y<1.5$.  Therefore, corrections to eq.~(\ref{magic}) are at most a few
percent larger than the deviation of $R(y)$ from unity.

$|f^{(D\to K^*)}|$ has already been determined.  $|f^{(D\to\rho)}|$ may be
obtainable in the future, for example from experiments at $B$ factories, where
improvements in particle identification help reduce the background from the
Cabibbo allowed decay.  The measurement~\cite{aitala2} $
Br(D\to\rho^0\,\bar\ell\,\nu_\ell)/ Br(D\to\bar
K^{*0}\,\bar\ell\,\nu_\ell) = 0.047\pm0.013$ already suggests that
$|f^{(D\to\rho)}/f^{(D\to K^*)}|$ is close to unity.  Assuming $SU(3)$ symmetry
for the form factors, but keeping the explicit $m_V$-dependence in $S^{(D\to
V)}(y)$ and in the limits of the $y$ integration, the measured form factors in
eq.~(19) imply $ Br(D\to\rho^0\,\bar\ell\,\nu_\ell)/
Br(D\to\bar K^{*0}\,\bar\ell\,\nu_\ell) = 0.044$.

$S^{(B\to K^*)}$ is obtainable from experimental data on $B\to
K^*\,\nu_\ell\,\bar\nu_\ell$ or $B\to K^*\ell\,\bar\ell$.  While the former
process is very clean theoretically, it is very difficult experimentally.  A
more realistic goal is to use $B\to K^*\ell\,\bar\ell$, since CDF expects to
observe $400-1100$ events in the Tevatron run II (if the branching ratio is in
the standard model range).  There are some uncertainties associated
with long distance nonperturbative strong interaction physics in this
extraction of $S^{(B\to K^*)}(y)$. But on average over $1 < y < 1.5$ this is
probably less than a 10\% effect~\cite{ligeti2}.  Consequently a determination
of $|V_{ub}|$
using this method with a theoretical uncertainty
 less than 10\% seems feasible. Like the situation with
$|V_{cb}|$ other methods, for example from the inclusive hadronic mass
distribution~\cite{falk3}
or from predictions for the form factors from lattice QCD~\cite{flynn} will be
necessary to have
confidence that the theoretical uncertainty is indeed this small.

\end{document}